\documentclass[11pt,a4paper]{article}
\usepackage{pdflscape}
\usepackage{authblk,graphicx,rotating,amssymb,amsfonts,amsmath,color,rotating,multirow,array,setspace,pdfpages}
\usepackage[margin=2.5cm]{geometry}
\usepackage[backend=biber,style=authoryear,natbib=true,uniquename=false]{biblatex}
\DeclareLanguageMapping{english}{english-apa}
\usepackage{orcidlink}
\title{The dynamics of diversity on corporate boards}

\author[a,b,d]{Matthias Raddant\,\orcidlink{0000-0001-6157-7733}}
\author[c,d]{Fariba Karimi\,\orcidlink{0000-0002-0037-2475}}

\affil[a]{Graz University of Technology, Institute of Software Technology, Inffeldgasse 16b/II, 8010 Graz, Austria, raddant@tugraz.at}
\affil[b]{University for Continuing Education Krems, Department for Knowledge and Communication Management, Dr.-Karl-Dorrek-Stra{\ss}e 30,
3500 Krems, Austria}
\affil[c]{Graz University of Technology, Institute of Interactive Systems and Data Science, Sandgasse 36/III, 8010 Graz, Austria}
\affil[d]{Complexity Science Hub Vienna, Josefst{\"a}dter Stra{\ss}e 39, 1080 Vienna, Austria}

\date{}                     %% if you don't need date to appear
\begin{document}
\onehalfspacing

  \maketitle
  
  \begin{abstract}
  Diversity in leadership positions, including corporate boards, is an important aspect of equality. It is important because it is the key to better decision-making and innovation, and above all, it paves the way for future generations to participate and shape our society. Many studies emphasize the importance of the visibility of role models and the effect that connectivity has on the success of minorities in leadership. However, the connectivity of firms, the dynamics of the adoption of minorities into leadership positions, and the long-term effects in terms of group dynamics and visibility are not well understood. Here, we present a model that shows how these effects work together in a dynamic model that is calibrated with empirical data of firm and board networks. We show that homophily -- the appointment of minorities is influenced by the presence of minorities in a board and its neighboring entities -- is an important effect shaping the trajectory towards equality. We further show how perception biases and feedback related to the visibility of minority members influence the dynamic. We find that reaching equality can be sped up or slowed down depending on the distribution of minorities in central firms. These insights bear significant implications for policy-making geared towards fostering equality and diversity within corporate boards.

\noindent \emph{keywords:} corporate boards, executive gender, diversity, firm networks
  \end{abstract}

\section{Introduction}

Corporate boards shape the governance and long-term strategy of a firm. Their decisions have lasting consequences for employees, shareholders, customers, and society as a whole. While the general composition of the board has been the target of numerous studies, diversity on corporate boards, and especially its dynamics, are not well understood, even though this would be paramount for devising actions to overcome deficiencies in diversity. In this paper, we present a diffusion model that describes important aspects of these dynamics. In particular, we illustrate the effect of homophily in the process of appointing new board members and its effect on the centrality and perception of minorities.

Women are still underrepresented in the corporate boards in most countries, even though there is noticeable regional variation (see \cite{imf_fem} for the EU, \cite{fem_us} for the US). In 2023, 35 percent of corporate board seats in Europe's large corporations were held by women (non-executive directors). The numbers are lower when focusing on the share of female CEOs. 8.4 percent of CEOs in the EU are women. In the US, for the first time ever, the share of female CEOs among Fortune 500 companies exceeded 10 percent in 2023.\footnote{See Fortune, June 5th 2023, Women {CEOs} run 10\% of Fortune 500 companies and \cite{EIGE}.}

Diversity has a positive effect on many dimensions of the firm's activities.
For example, from studies in management science, we know that diversity has a positive influence on decision-making under risk, on creating innovations, and on recruiting new talents  \citep[see, e.g.,][]{glassbreaking,div_perf,board_diverse}. Diversity has been identified as a predominantly positive influence on the efficiency of teams, in particular in top management \citep{yinyang,post_gender}.

It is more difficult to identify effects with respect to financial effects of diversity. For example, \citet{ahdit} find that the increase in female board members might have led to less experienced boards and thus decreased performance,  \citet{green_fem} and \citet{camp}, however, find the opposite effect. From an investor's point of view, gender does not seem to play a role \citep{brink}. With respect to earnings volatility and firm performance \citet{faccio} and \citet{jpnfem}  find tendentially positive effects from board diversity and female CEOs.

Hence, while the under-representation of women, or minorities in general, on boards might not be much of an issue in the short-run, there is evidence that it influences monitoring \citep{adams_fem}, the pace of innovation within the company \citep{femnorw,fem_rep,minor_ceo,ostergaard2011does}, as well as the setting of societal norms and practices within the corporation and outside of it \citep{matsa,kirsch21,dezso,keller_gap}.

A useful measure of the long-run performance of the company is its ability to devise strategies  with respect to sustainability, which can typically not be achieved without financial soundness. 
An increasing amount of studies hints at the positive influences of diversity on sustainability \citep{board_sust,ecc_sus_man,khan_sus_accounting}.
The appointments of female board members may, therefore, also serve as an indicator of social responsibility in general and is likely to influence the career paths of future generations of female executives \citep{bilimoria,farrell}. When it comes to further dimensions of diversity there is some literature about the (tendentially positive) influence of LGBTQ-related policies on firm performance and firm attractiveness as an employer \citep{lgbt2,lgbtqf}. 

The ongoing process of new appointments of board members takes place in a networked environment, where managerial practices, governance, and appointment decisions are influenced not only by the composition of the members but also by what is observed in other corporations \citep[see also][]{Borgatti03,branson}. The inter-organizational networks within which these managerial practices diffuse are manyfold: they stem from ownership relationships, supply chains, contacts to customers, and ties between the board rooms of corporations \citep[see also][]{neteffects,cai,HW03}. Other determinants for the likelihood of hiring women can be found by examining board members' personal lives, in particular, family status and aspects of their formative years \citep{cronq_fem,gender_gap}.

Despite the documented diffusion of practices via these different networks, there are only few instances where the influence of networks on diversity has been explored. The most notable exceptions are probably the small number of recent papers in management science that focus on the importance of networks and similarity of board members for the appointment process. \citet{cai_who} find that most new board members have ties to incumbent board members. While these ties increase the chances for being appointed, \citet{lalanne} finds that homophily between existing and new board members with respect to socio-demographic characteristics might hinder the creation of diversity on boards. Using different methodology, related observations were previously made by \citet{zhu_ingroup}, who found that minorities might be subject of subtle discrimination with respect to board appointments, and that they are also less likely to serve in leading positions once they are appointed \citep{field_glass}. \citet{west_stern} found that while boards might appoint directors who are different in one dimension, these then have to be more similar to existing board members in other dimensions, \citep[see also]{ingroup}. Similar effects of homophily can be found with respect to the appointment of foreign board members \citep{barrios_hom}.

%Studies have also shown that women often tend to connect to other women more than men to men, presumably due to in-group support \citep[see also][]{ingroup}. Additionally, the group of female board members, of course, forms a minority in an otherwise male-dominated network, which makes homophily one factor for the appointment of female board members \citep[see also][]{birds}.

Hence, while some studies have analyzed panel data and network effects on equality among executives, it is difficult to draw conclusions on the dynamics of diversity in the medium to long run, because we lack the modeling of the network itself and the dynamics processes related to it. In order to evaluate governance and policy interventions geared towards more diversity, it is therefore necessary to connect this literature to
the principles found in generative network models, because they offer the foundation for modeling the necessary transitions towards more diversity over time.

For example, the effects of homophily within groups and the implications for connectivity, centrality, and visibility of individuals have been analyzed analytically by \citet{eun} in extensions of the scale-free network model of \citet{Barabasi99}. These models produce networks with a power-law degree distribution. However, when different groups are introduced and when these groups are assigned different levels of homophily, the tail exponents of the degree distributions for the groups will differ as a result of this homophilic behavior.
This has implications for the visibility of a group, which depends dis-proportionally on nodes in the top ranks of the distribution. 
%Additionally, the perception of minorities varies as a function of such structural differences \citep{eun}. 
While such models have been applied to data sets of scientific citations and have been successful in pointing out mechanisms that lead to inequality in academic careers \citep{econ_colab,cite_neuro}, an application to corporate boards and executives is still missing. 

This paper, therefore, presents a network-based diffusion model that takes into account the empirical findings on board member appointments and connects them to generative network models. We simulate the appointments of new board members over a time span of several decades. In particular, we are interested in the effects of homophily in the process of new appointments with respect to gender (and in general minorities), because homophily has been identified as a relevant aspect in the analysis of corporate board composition and in the analysis of diversity in social networks in general \citep{green_fem,jpnfem,bram}. Our results show that homophily has a significant influence on the representation of minorities in terms of network centrality, as well as on the perception of their group size.  We further show that bias in the perception of minorities is a plausible feedback mechanism that influences the speed of the development toward equality.

%In the following we give a brief overview of network model and how they can describe homophily in section \ref{sec:models}. Section \ref{sec:jpn} summarizes the empirical findings for corporate boards in Japan, which are the foundation for our model, which is described in section \ref{sec:sim}. We then discuss the results for the case of gender as well as for the more general case of a minority in section \ref{sec:results}. Section \ref{sec:conc} concludes. \Fariba{I suggest to eliminate this paragraph}

\section{Network models and diversity}\label{sec:models}

We will start with a brief look at a class of network models that constitute the theoretical backbone of our analysis.
The dominant class of generative models to describe most large-scale social and economic networks is the scale-free  model \citep{Barabasi99}.
The algorithm behind this model rests on two mechanisms: growth and preferential attachment. This leads to networks with a  power-law degree distribution
\begin{equation}
p_k \propto k^{-3}
\end{equation}
 for large degree $k$. An implication of this is that a node $i$ has an age $t_i$ defined by the time the node was added. Nodes increase their connectivity over time as
\begin{equation}
k_i(t) = \left(\frac{t}{t_i} \right)^{1/2} \; .
\end{equation}
Apart from age, nodes are homogeneous, there is no chance to `overtake' other nodes in terms of degree development. This can only be changed if we add additional parameters to the attachment mechanism \citep[see also][]{agingofsites}.
If we assume fitness levels $\eta_i$ for nodes, the probability of a new node to connect to node $i$ becomes
\begin{equation}
 \Pi_i = \frac{\eta_i k_i}{\sum_j \eta_j k_j} \; .
\end{equation}
This opens the opportunity for young nodes with high fitness to attract links with high rate and to `overtake' older nodes \citep{bianconi}. 

  \begin{figure}[tb]
\begin{center}
\includegraphics[width=0.3\textwidth, trim = 0 85 0 85 , clip=true]{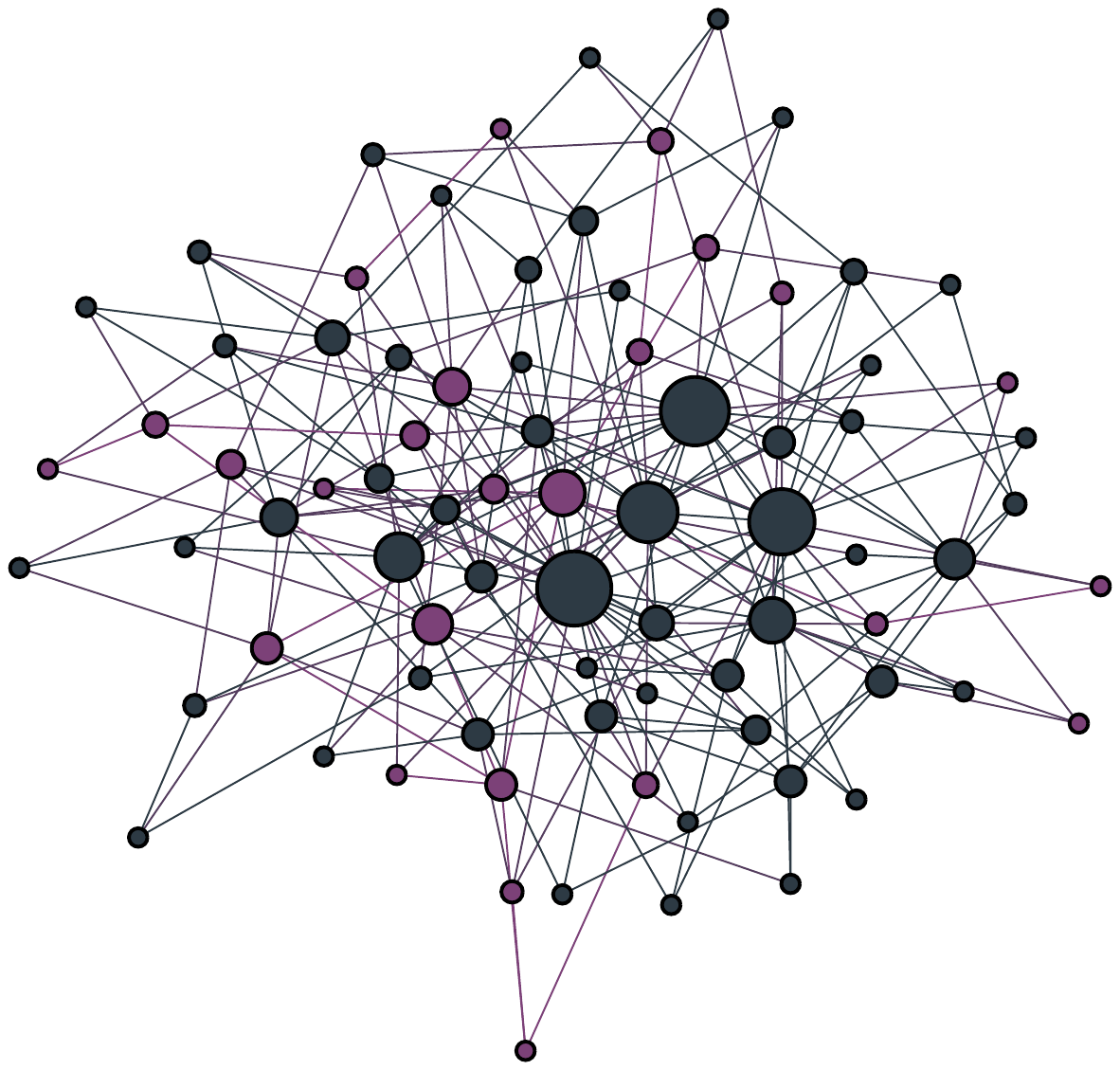}
    \includegraphics[width=0.3\textwidth, trim = 0 85 0 85, clip=true]{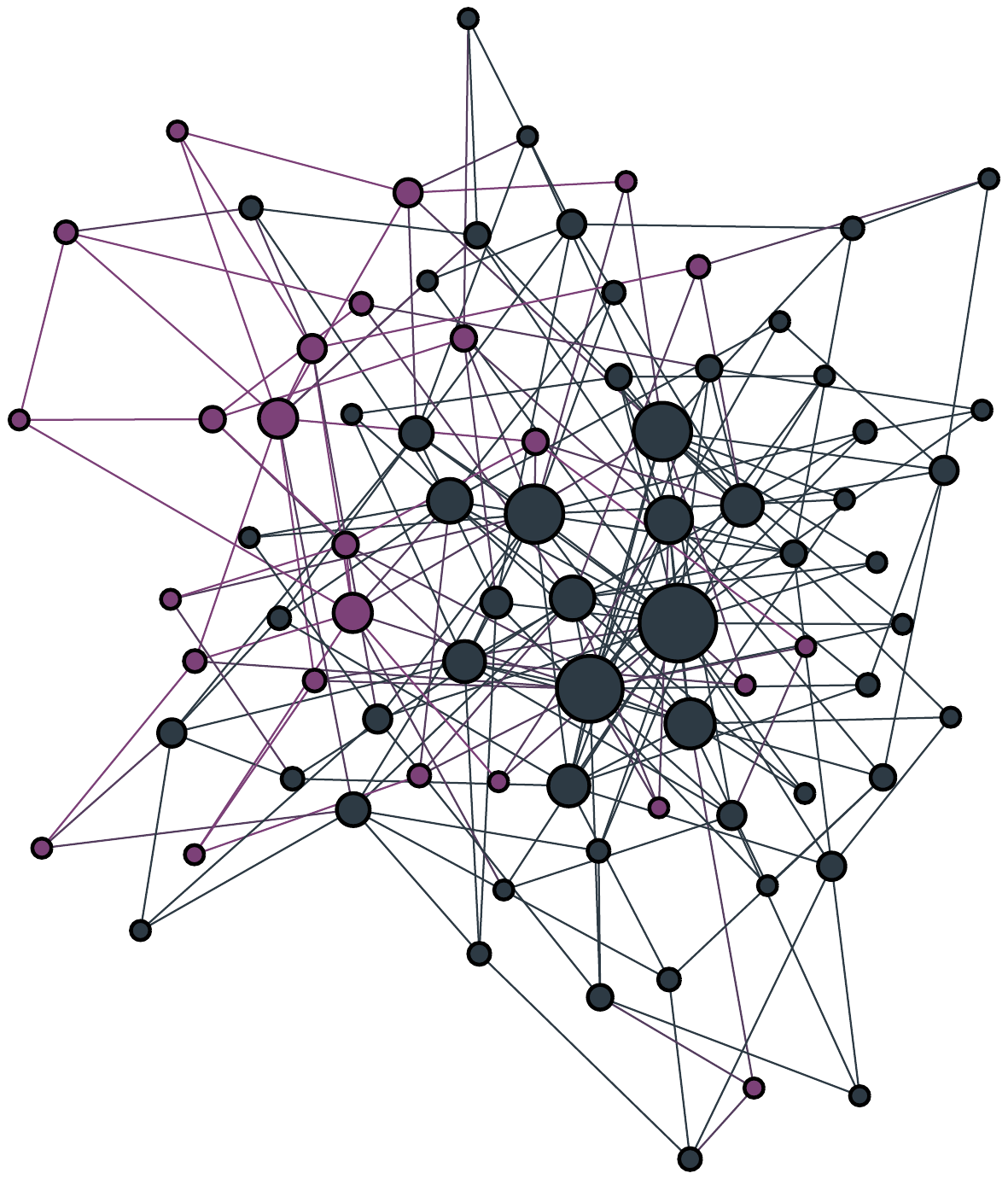}
    \includegraphics[width=0.3\textwidth, trim = 0 85 0 85 , clip=true]{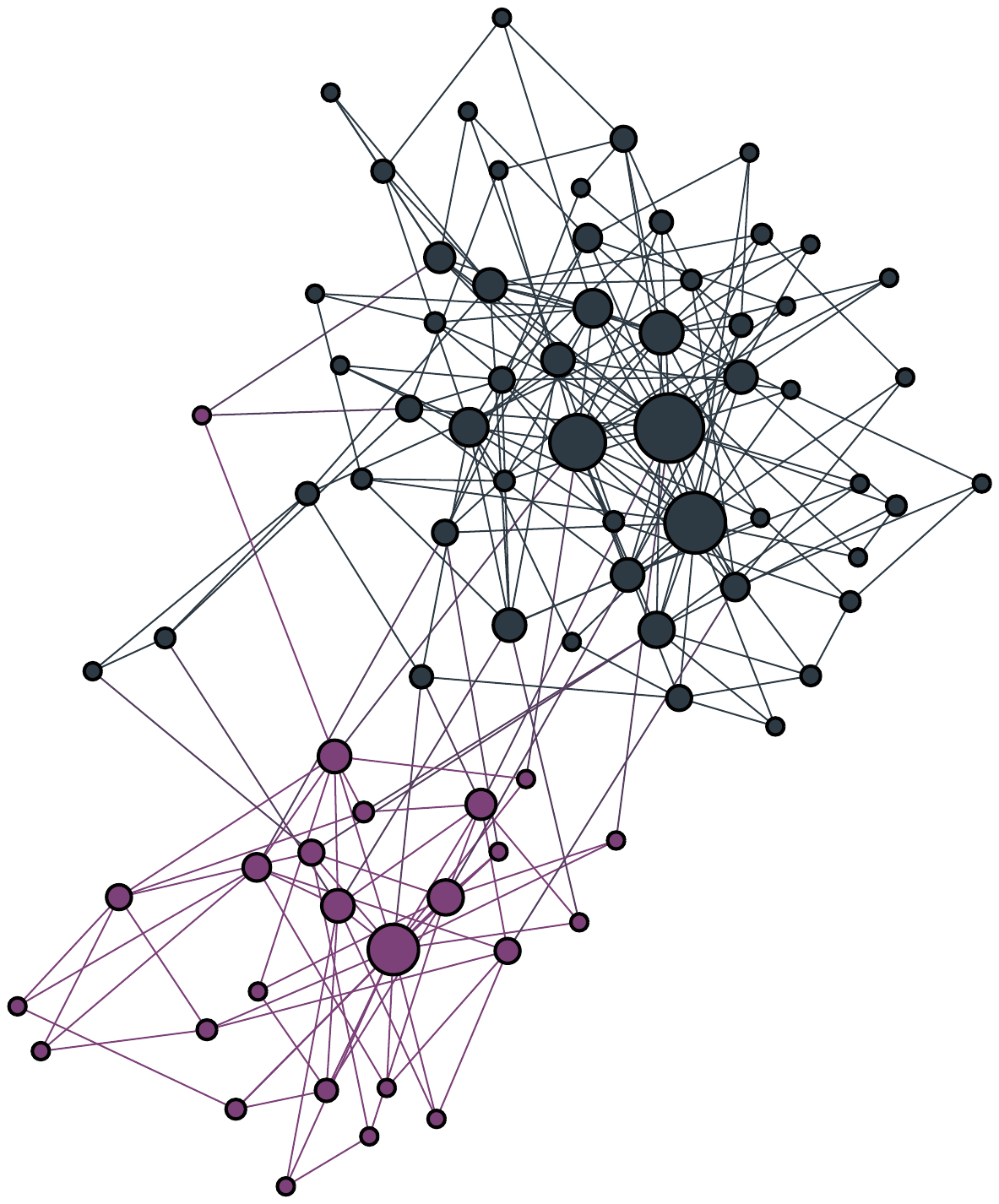}
        \includegraphics[width=0.3\textwidth, trim = 0 85 0 85 , clip=true]{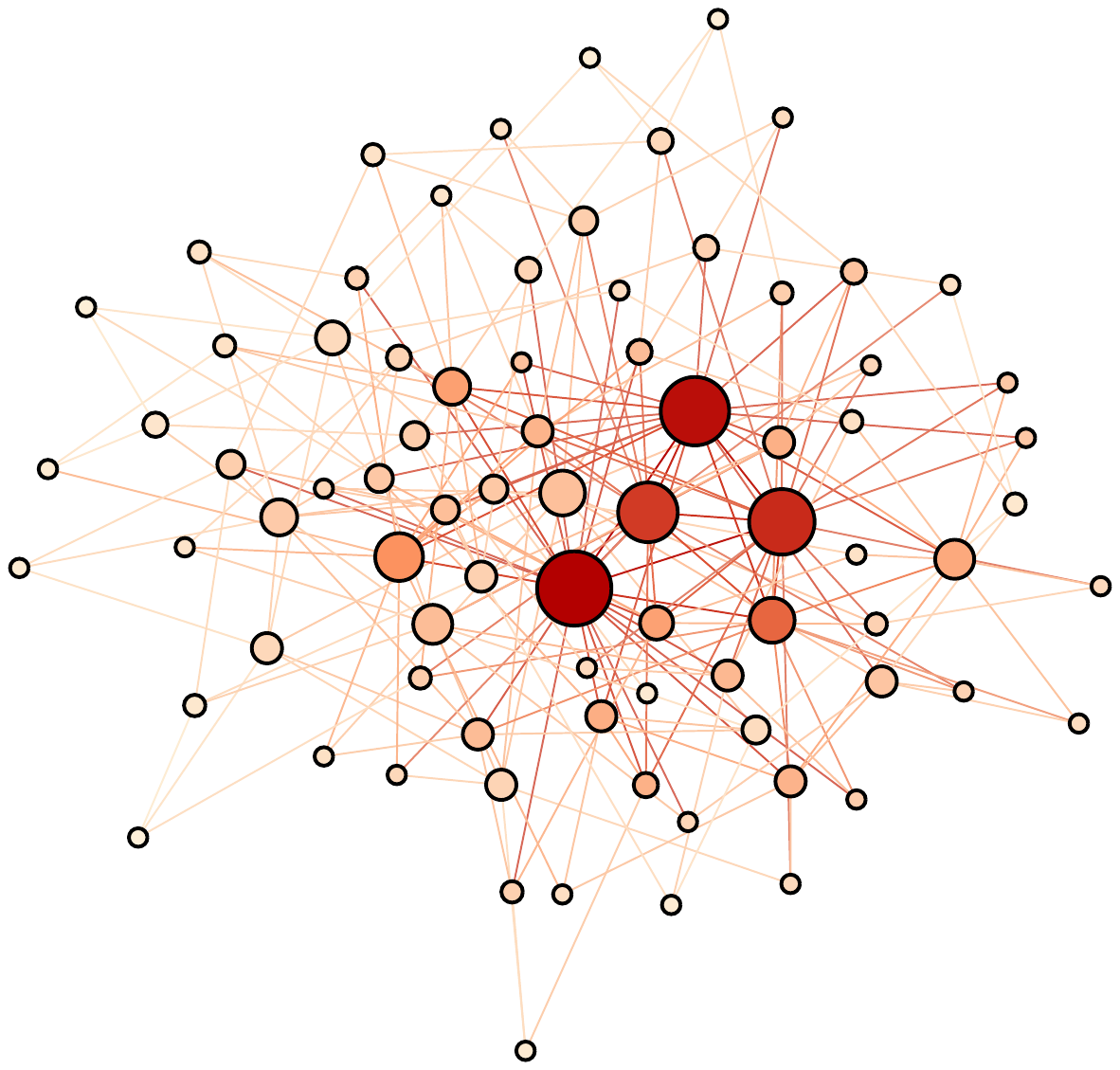}
    \includegraphics[width=0.3\textwidth, trim = 0 85 0 85 , clip=true]{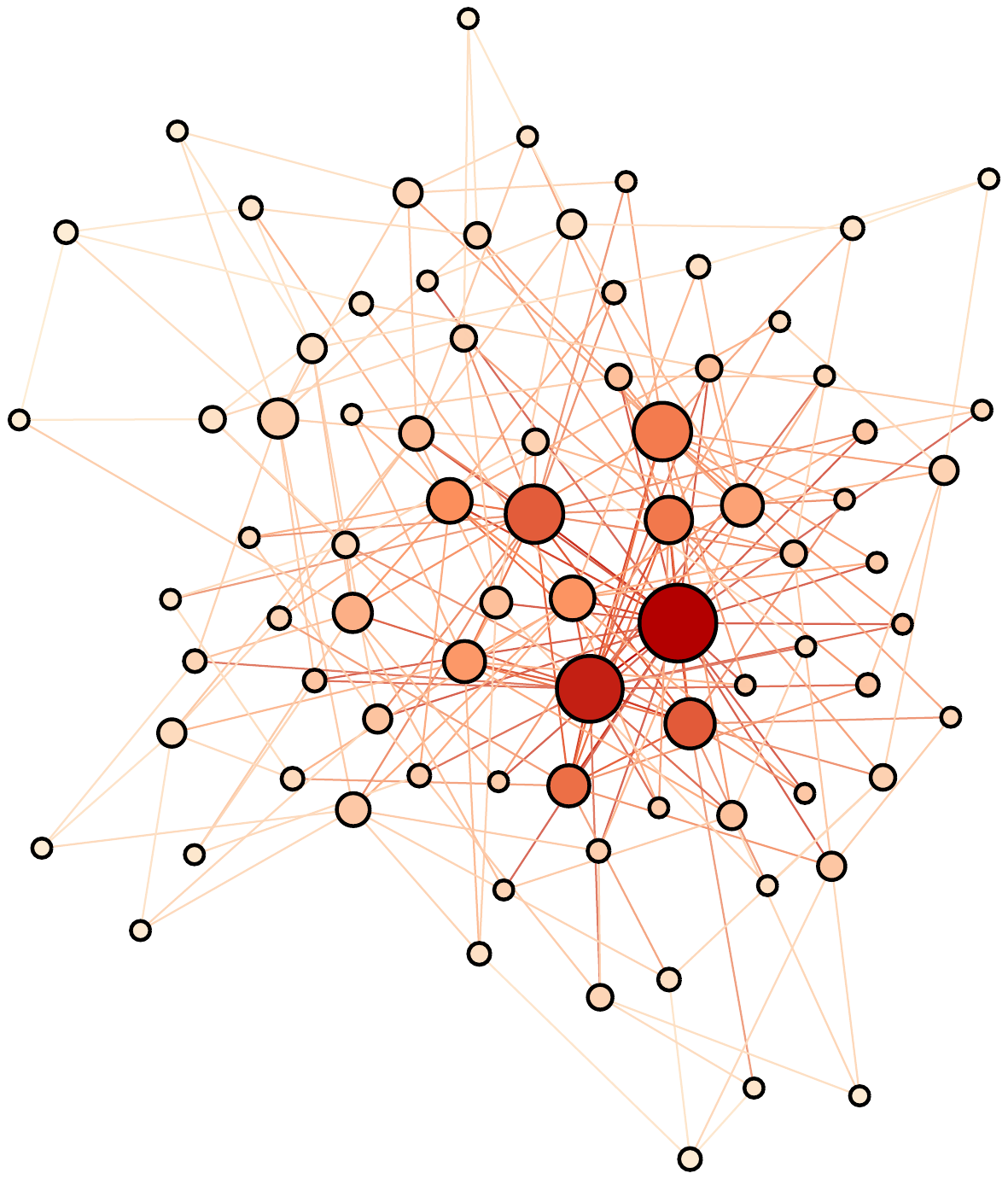}
    \includegraphics[width=0.3\textwidth, trim = 0 85 0 85 , clip=true]{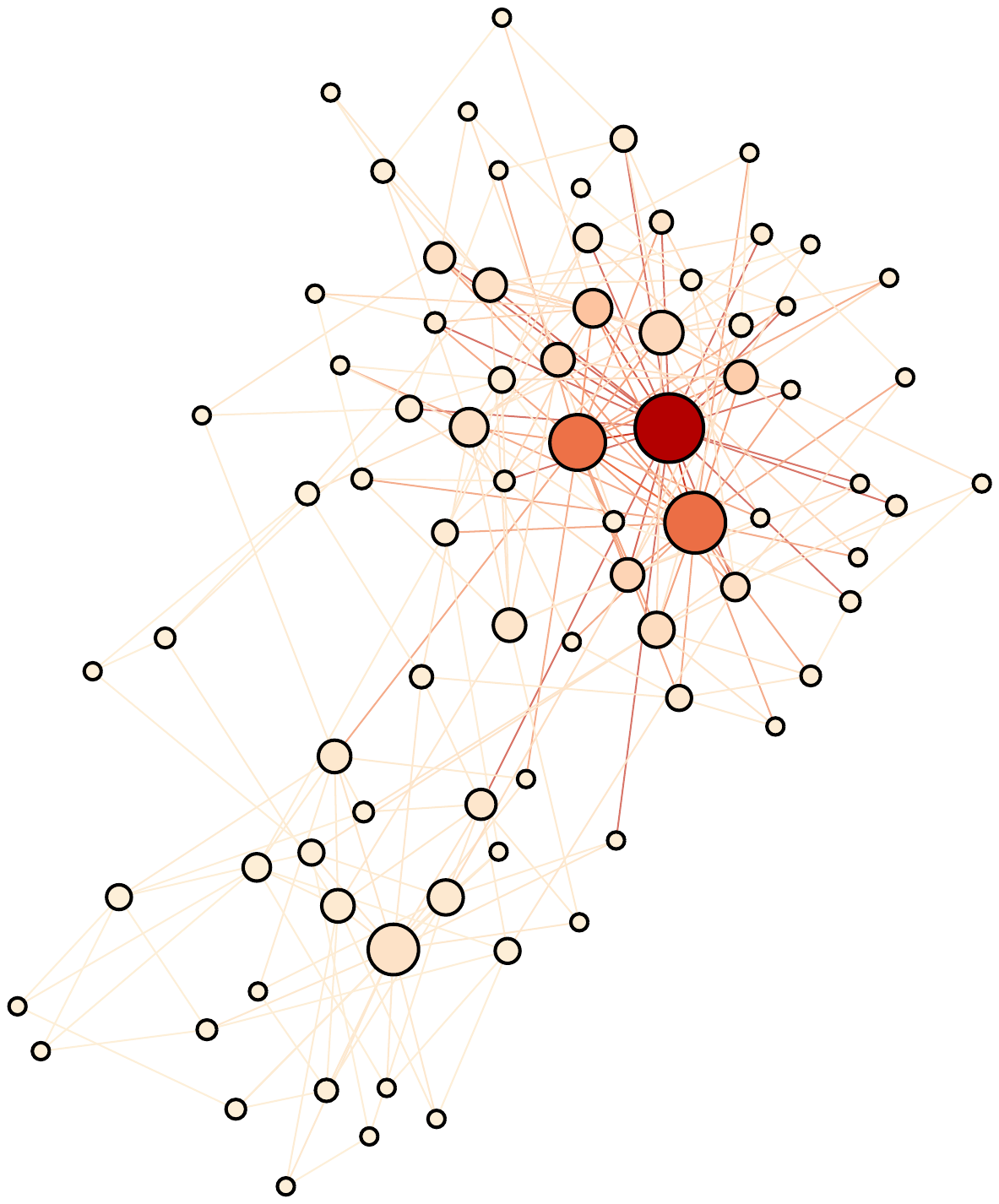}
  \end{center}
  \caption{Three examples for homophily networks, $h = 0.5, 0.7, 0.9$
  left to right, $N=80$, minority size 30\%. The node size is proportional to node degree. The top panels show the nodes colored according to group classification. The bottom panels show the same networks with nodes colored according to eigenvector centrality. Note that the minority group does not contain any highly central node for $h\geq 0.7$.}\label{fig:hnets}
\end{figure}

The above formulation can also be used to define group-specific attachment rules. This leads to the model for group homophily \citep{karimirank}.
We define $f_a$ as the fraction of nodes in group $a$ and $f_b = 1 - f_a$ for group $b$.
Homophily $h$ can range from 0 (perfect heterophily) to 1 (perfect homophily). When we assume symmetric and complementary homophily, $h = h_{aa} = h_{bb}$ and $h_{ab} = h_{ab} = 1-h$
the probability of node $j$ to connect to $i$ becomes
\begin{equation}
 \Pi_i = \frac{h_{ij} k_{i}}{\sum_l h_{lj} k_l} \; .
\end{equation}

The expected degree then depends on homophily and the relative group sizes.
This is mirrored by the exponents of the degree distributions for nodes in group $a$ and $b$, %\citep[see also][]{karimirank}
\begin{equation}
p_{a} \propto k^{\gamma_a} \; \;
\end{equation}
\begin{equation}\nonumber
  p_{b} \propto k^{\gamma_b}  \;.
\end{equation}
Homophily thus has implications for the centrality of nodes.
Nodes in the group that constitutes the minority have a lower average degree. They will likely not be important hubs in the network and will rank low in eigenvector centrality, see figure \ref{fig:hnets} for an illustration.
Homophily also influences neighborhood-based estimates of group size.
Assume that each node has a local perception of group sizes, defined as the share of $a$ and $b$ nodes among its direct neighbors. The local perception of the share of $a$ nodes by $i$, $s^a_i$, can then be expressed as

\begin{equation}
s^a_i = \frac{N_a^i}{N_a^i+N_b^i} \; .
\end{equation}
There are different ways to express biases in these perceptions. Let us write the true group size as
\begin{equation}
s^{a} = \frac{N_a}{N_a+N_b} \; .
\end{equation} 
Then we can express the deviation between the average of the perception of $a$ nodes about the group $a$ and the true group size as
\begin{equation}\label{eq:perc}
\Delta s^a = \frac{ \langle s^a_{i\in a}\rangle}{s^a} \; .
\end{equation}
 One can show that in homophilic networks ($h > 0.5$) the minority
overestimates its own size, while the majority underestimates the size of the minority \citep{eun}.

\section{Empirical foundations for modeling board appointments}\label{sec:jpn}

Our modeling approach in this paper is based on previous empirical research on the ties between corporations and their board members, as well as their survival over time. We have analyzed 4,505 publicly listed firms and 95,192 board members in Japan from 2004--2013.
We found that the `survival' rate for board members lies around 85\% each year. The replacement of board members can -- on the macroscopic level --  be described as a stochastic process that is influences by factors like firm survival, the number of mandates of board members, and whether the board member acts as an auditor or outside member. To a lesser extend other factors such as age and firm size play a role. The influence of gender and firm profitability on the survival of board members in the short-term is negligible \citep[for details see][]{jpnboards}.
 
From 2004-2013, the share of female board members in Japan slowly rose to about 2\%.
%Some differences exist between sectors, the share of female board members is highest in insurances and services, lowest in mining, iron and steel, marine transportation, and construction. 
We observe that female board members initially were more likely to be hired at smaller firms \citep[see also][]{saito, MORIKAWA20161}. We also find evidence for a catching-up of female board members in terms of centrality. Most importantly, we find homophily with respect to the distribution of female board members within corporate networks \citep[for details see][]{jpnfem}.

%To illustrate this finding we have revisited the data set to calculate conditional probabilities for the hiring of female board members\footnote{For this calculation the sample is reduced to 2,286 firms for which financial information was available and that were part of the giant component of the network in the year when hirings took place.}.  We observe a total of 27,805 hirings. In 502 cases a women was hired (1.81\%).
% We calculate probabilities conditioning the hiring of a female board member on observing existing female board members on the own board, through ties in the board network, and through ties in the ownership network. We find that the following conditional probabilities: 1.27\% (no women present), 2.23\% (1 women present in board network), 2.46\% (3 women in ownership network), 3.70\%  (1 women on own board), 5.25\% (1 women on own board, 1 through board network), 5.56\% (2 women on own board).

This homophily also partly explains the appointments of new board members.
We found that the appointment of women was more likely when there was already a women present on the board of a company or (to a lesser extend) when a woman was present on the board of a firm to which connections exist via ownership ties or corporate board interlocks.
 
Hence, while this data set - due to the low overall number of female board members - does not allow for a modeling of endogenous growth of groups and their networks, it is a good foundation for modeling the dynamics of board members' replacement in terms of a diffusion model that describes the likelihood for female board members to be hired for specific vacant seats on corporate boards.\footnote{Other recent studies on corporate boards in Japan include \citet{tanaka_fem}, \citet{kubo_jpn} and \citet{tokenism}. For an overview on ownership structures in Japan see \citet{jpn_owners}.}

\section{A model for diversity}\label{sec:sim}

\begin{figure}[tb]
\begin{center}
    \includegraphics[width=\textwidth, trim = 40 475 5 150, clip=true]{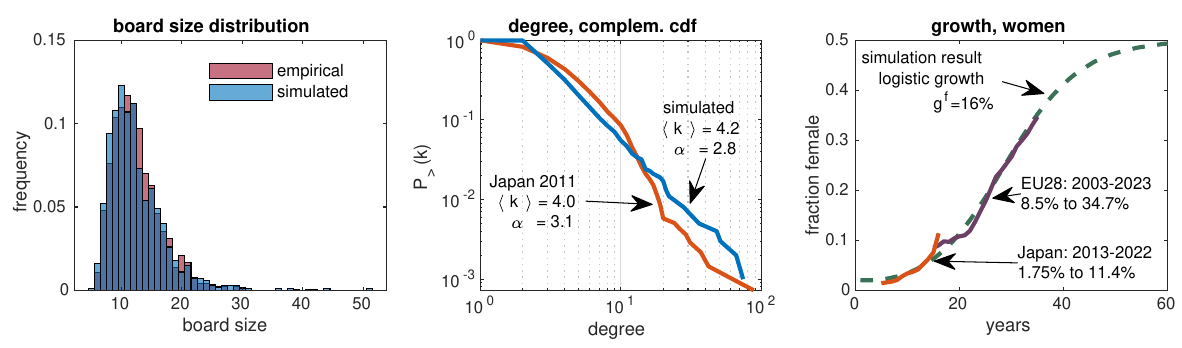}
  \end{center}
  \caption{Model calibration overview: the left panel shows the board size distribution for 2011 compared with one realization from the model. The top middle panel shows this comparison for the degree distribution. The right panel shows the fraction of female board members from empirical data together with that from the model (assuming logistic growth in the number of newly appointed female board members).
  }\label{fig:calib}
\end{figure}

\subsection{Corporate boards and firm networks}

To analyze the dynamics of diversity on corporate boards, we define a diffusion model which shows the effects of homophily in the process of appointing new board members. A key ingredient of this model is that we assume that appointments of women are more likely at empty seats of those boards, which have other women already present in their direct neighborhood.  We show the effects on the representation of women in firms with respect to centrality, as well as the effect on the perception (visibility). Finally, we study how perception can act as a feedback mechanism for the adjustment speed of the system towards equality.
 
We simulate the composition of the boards of 1,000 firms in time steps of 1 year. The board size distribution, the firm network degree distribution and the board member turnover per year is calibrated to match the observed sample of corporates in Japan in the early 2010s, see the left and middle panel of figure \ref{fig:calib}. The board sizes follow a log-normal distribution with a mean of 12.5 and variance a of 20.6.

The firms (and their respective boards) are connected by fixed scale-free networks that represent an approximation of the merger of observed board and ownership ties, with an average degree and power-law exponent calibrated to resemble empirical data.\footnote{We note that -- as most networks with ties subject to social and legal constraints -- the empirically observed networks do not strictly follow a power-law \citep[see also][]{broido_pl}, but can without loss of generality for the simulation results be reasonably well approximated by this model.} In the generation of the firm networks, we assume that firm size is synonymous with degree, and thus the ordering of the degree sequence aligns with the ordering of the board sizes. This is in line with our empirical findings and is, of course, partly an effect of legal guidelines, which necessitate larger boards at larger firms, and also mirrors the greater involvement in cross-share holdings and thus ownership ties of larger companies.

For the initial assignment of board members to firms, we consider two scenarios. The first scenario is to assume that female board members are assigned to firms in an unbiased way, i.e., the share of women is independent of other characteristics, in particular firm size. However, studies have shown that women often serve on the boards of smaller firms, which is likely to have implications for their visibility and might influence the hiring of new board members in the future. Considering this, we also investigate an alternative scenario where the probability for each seat $i$ on a board to be filled by a woman is in an inverse relationship with the degree of the corresponding firm $j$. We define
\begin{equation}\label{eq:size}
prob(v_i^0 = f) \; \propto \;  1 + \gamma \; \frac{\langle k \rangle - k_j}{k_j} \; ,
\end{equation}
%\Fariba{in above equation, I guess that we mean $k_i$ and not $k_j$}
which describes a function where the probability slowly decays with increasing degree $k$.\footnote{$\gamma = 0.8$ is a parameter that describes the intensity of this relationship. The generated networks satisfy $k\geq 1$.} This specification allows us to assess how the initial assignment of seats influences the dynamics of board member appointments. 

\subsection{Appointment of new board members}

In line with empirical data, we assume that 15\% of the board members retire each year, and that new board members are appointed. We start with an overall share of 2\% female board members. We assume that the share of female board members that are appointed each year slowly converges towards 50\%, following a discrete logistic function. The growth rate of this logistic function has been calibrated such that the simulation results roughly match empirical data from Japan and the EU\footnote{Data from \citet{EIGE} and \citet{jpn_official}. Note that in our calibration of the growth rate we disregarded shifts due to policy and regulatory changes, in particular in 2022 in Japan and around 2009 in the EU.}, see the right panel of figure \ref{fig:calib}. Hence, denoting the number of retiring board members as $N^r$, we can write the number of appointed female board members in year $t+1$ as
\begin{equation}\label{eq:growth}
x(t+1) = x(t) + g^{f} \; x(t) \frac{2 \big( 1-x(t) \big) }{N^r},
\end{equation}
where $g^f=0.16$ is the growth rate.

\begin{figure}[tb]
\begin{center}
    \includegraphics[width=0.95\textwidth ]{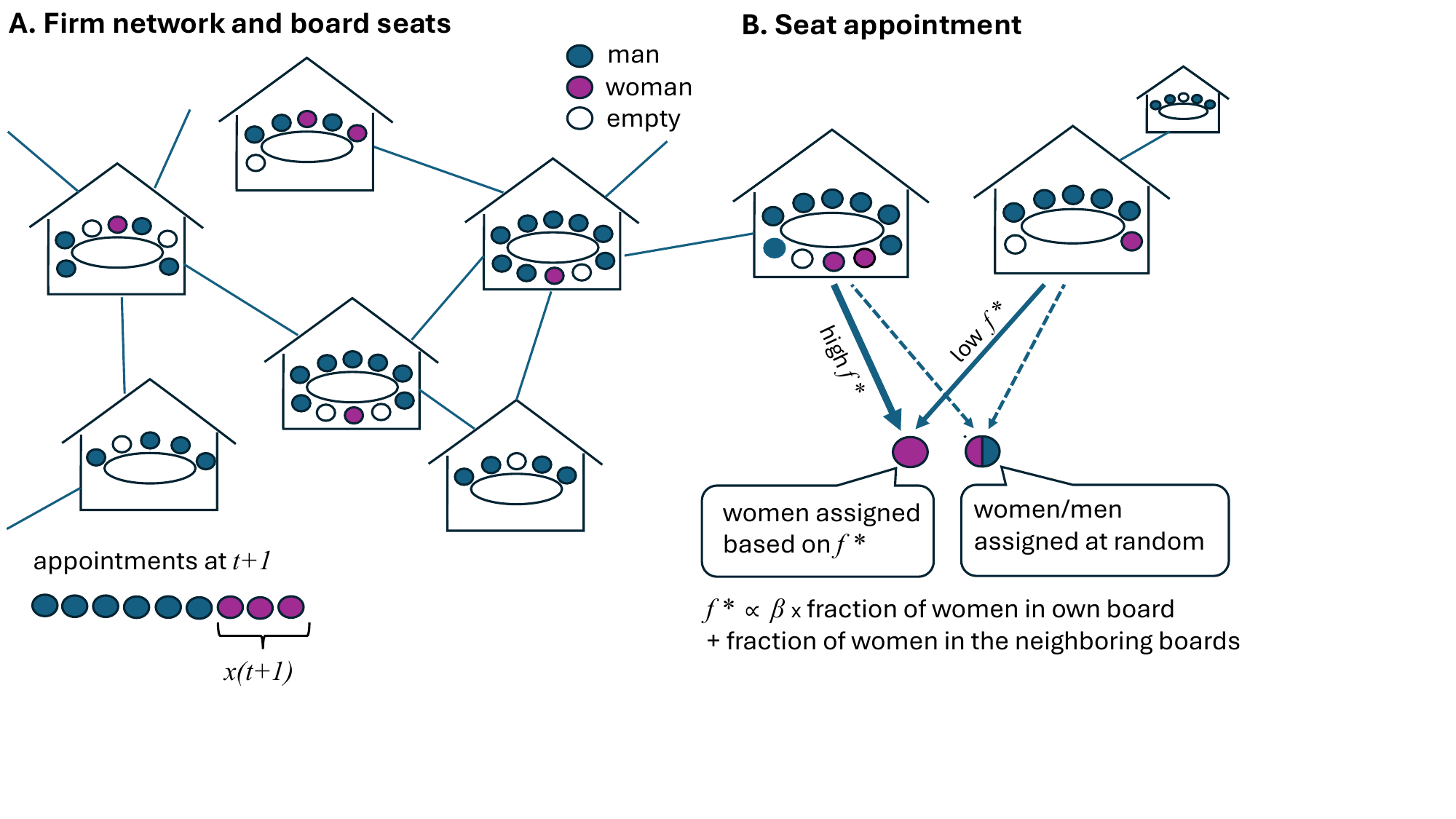}
  \end{center}
  \caption{Overview of assignment dynamics: boards with different sizes are connected by a static network (panel A). For the appointment of new board members, we assume that for a fraction of women $\lambda$ seats are assigned based on $f^*$. For the remaining women and all men seats are assigned at random (panel B).
  }\label{fig:model}
\end{figure}

While this means that the shares of men and women who fill the vacant seats is exogenously given, we assume that in the decision \emph{where} a women will be hired homophily plays a role. In particular, we assume that at each time step a certain share $\lambda(t)$ of the vacant board memberships for women is chosen according to homophilic principles.
With respect to this share, we analyze two different scenarios. 
Since there is some empirical evidence that the degree of homophily in social networks changes dependent on group size \citep[see][]{birds} we start by modeling homophilic behavior that diminishes when group sizes equalize. We can define the current group size (the share of women or the minority) as
\begin{equation}
 y(t) = \frac{N^f(t)}{N} \; 
\end{equation}
 and define the homophily variable such that 
 \begin{equation}\label{eq:lambda}
 \lambda(t) = min \; \left( \; \bar{\lambda} \; , \;  1 - \left( 1+exp \big( -g^{\lambda} \big( y(t)-y^m  ) \right) \right)^{-1} \; .
 \end{equation}
Thus, the share of seats assigned with homophilic behavior follows a (reversed) continuous logistic function (with $g^\lambda = 20$ and the mid-point $y^m = 0.16$). This means that the homophily will go to zero once the group sizes equalize. Since it is unrealistic that homophily plays a role in each hiring decision, we impose an upper bound $\bar{\lambda}=0.9$, which means that at least 10\% of vacant seats for women are always assigned randomly.\footnote{The exact value of $\bar{\lambda}$ does not play a decisive role, since also the `less desirable' seats of the homophilic assignment are rather random. The upper bound also ensures that the growth of the minority can never outpace the number of available seats. An alternative to ensure this would be a modification to equation \ref{eq:fi} that ensures positive $f_i^*$.}

For the homophilic seat assignment we assume that the likelihood of a particular vacant seat to be filled by a woman depends on the number of existing women in the direct neighborhood of the vacant seat at firm $i$. Our empirical data suggests that the probability depends on the share of female board members on the board  of firm $i$, $N_i^{f,B}/N_i^B$ and, to a lesser extent, on the share of existing female board members on the boards of connected firms, $N_i^{f,C}/N_i^C$. Hence,
\begin{equation}\label{eq:fi}
prob(v_i = f) \propto f_i^* = \frac{N_i^{f,C}}{N_i^C} + \beta \; \frac{N_i^{f,B}}{N_i^B} \; ,
\end{equation}
where the parameter $\beta=2.5$ weights the two influences. Probabilities are obtained by normalizing (\ref{eq:fi}) by $\sum_i f_i^*$. An overview of these mechanisms is provided in figure \ref{fig:model}.

The above formulation assumes that the decline in homophily is caused by a change in behavior that depends on group size. Since there is only little previous research on this effect, we also analyze the possibility that the change is caused by structural constraints of the network, while the behavior stays unchanged.
This means that we can alternatively assume that the share of seats assigned based on homophily stays fixed throughout the simulation. In this case 
\begin{equation}
\lambda(t) = \bar{\lambda} = 0.9 \; .
\end{equation}
In this model, aligned with empirical findings, we assume that no homophilic behavior is present in the hiring process for men (the majority).

\subsection{Model variations}

\subsubsection*{Dimensions of diversity} 

Instead of interpreting the groups in this model in terms of gender we can of course interpret them in any other sense of diversity. In particular, by changing the final proportion of the group sizes to a setting where one group stays significantly smaller than the other, we can create a situation that characterizes a minority which group size is growing towards its actual share in the population.

\subsubsection*{Endogenous growth and visibility}

An important aspect of the modeling of the dynamics of diversity is of course the speed of this process. While the aspects of the model described above were to some degree based on empirical data, we cannot obtain an approximation for the determinants of the growth rate  in equation \ref{eq:growth}. Studies on the importance of role models and the feedback effects from representation in leading positions \citep[see, e.g.,][]{bilimoria} however hint that the visibility and perception of minorities have an important effect on the career paths of future leaders. This motivates to endogenize the growth in the number of new female board members by employing the formalism that describes the perception of female board members from equation \ref{eq:perc}.

In order to include the visibility of female board members into equation \ref{eq:growth}, we assume that the number of new female board members in each time step $t+1$ is influenced by the deviation of the perceived from the actual number of women, $s^f(t)$, at time step $t$.

This means that we can modify the growth equation to
\begin{equation}\label{eq:endo}
x(t+1) = \left( 1 + \Delta s^f(t) \right) \left( x(t) + g^{f} \; x(t) \frac{2 \big( 1-x(t) \big) }{N^r} \right) \; .
\end{equation}
Hence, we assume that the grow of the inflow of women to the boards can be accelerated or slowed down relative the baseline parametrization, depending on biases in the perception of the number of women, which itself are caused by differences in the centrality of women versus men.\footnote{A calibration of the effect described by equation \ref{eq:endo} could be achieved by introducing an additional parameter as a proportionality constant.}

\subsubsection*{Limitations} 

We note that while important aspects of our model are calibrated with empirical data, the model is not meant to provide a forecast with respect to 
the exact extent of gender equality in the future. It is meant to show effect sizes for different scenarios, which to our best knowledge, capture the magnitude of effects that have been observed in practice. 
One of the structural limitations of this model is that we look at static networks. While links in institutional settings are in fact more persistent than in social contexts \citep[see also][]{survey_persistance,kogut,persist}, board memberships do in principle endogenously change the network structure to some extent. Since our data set is insufficient to model these, and network re-wiring is not the scope of this analysis, these aspects are left for future research. 
We also note that the formalization of endogenous growth is an ad hoc assumption. Hence, while the relationship between observed group sizes and the rate of newly appointed women in equation \ref{eq:endo} is consistent from a modeling perspective, we do not have empirical data that would allow for a validation and more precise calibration of this effect.

Also, while this model emphasizes initial conditions and successive dynamics, it can be expected that additional effects come into play once group sizes change. Structural effects as well changes in choice will influence the outcome with respect to the final group sizes and their actual representation on corporate boards vs their theoretical limits.

\section{Results}\label{sec:results}

\subsection{Scenarios towards diversity in firms} 

In the following, we will analyze different scenarios and how they influence the distribution of board members in terms of centrality and visibility over time. 
In particular, we are interested in the role of homophily in the growth process. We investigate the interplay between the behavioral homogeneity in the hiring process and the resulting homophily in the firm network. Further, we analyze how the initial conditions with respect to the centrality of members of the minority impact visibility. Finally, we analyze the implications of feedback from the visibility on the dynamics of diversity.

We consider the following cases:
with respect to the \emph{initial assignment} of board seats we will begin by assuming that those are independent of firm centrality. Alternatively, we will assume that the minority is at the beginning assigned seats at less central (smaller firms) according to equation \ref{eq:size}.
For the \emph{homophily in hiring} we will investigate the case where this effect diminishes when group sizes equalize (see equation \ref{eq:lambda}), versus the case where the homophily in seat assignment is constant.
With respect to the final \emph{group size} we will analyze the case of an equalization of group sizes (the `gender' case) versus a minority that grows to $\frac{1}{6}$th of the total population.
Finally, we will replace our assumption about \emph{exogenous growth} (equation \ref{eq:growth}) with an endogenous growth where the speed (but not the final group size) receives feedback from group perception (see equation \ref{eq:endo}).

 As the outcome we observe the following variables:
the \emph{outcome homophily} in the firm network is defined as the correlation between the share of women (the minority) on the board and the share of them among the boards of directly neighboring firms in the network. We measure the deviation from an equal \emph{representation} of women (the minority) among the most central firms by calculating by how much women are over- or underrepresented in the most to least central firms. Note that since we do not model personal networks per se, we can regard the centrality of the firm as an approximation for the centrality of the board member itself. Finally, the \emph{perception} of the group of women (minority) board members is measured as the average share of them that each group observes on the boards of direct neighbors in the network. We report the perception relative to the actual group size.
All scenarios were computed with $1,000$ firms and $10,000$ simulation runs.

\subsection{Scenarios with exogenous growth}

\subsubsection*{Unbiased seat assignments and homophily in hiring as a function of size}

\begin{figure}[p]
\begin{center}
    \includegraphics[width=\textwidth, trim = 15 130 0 40, clip=true]{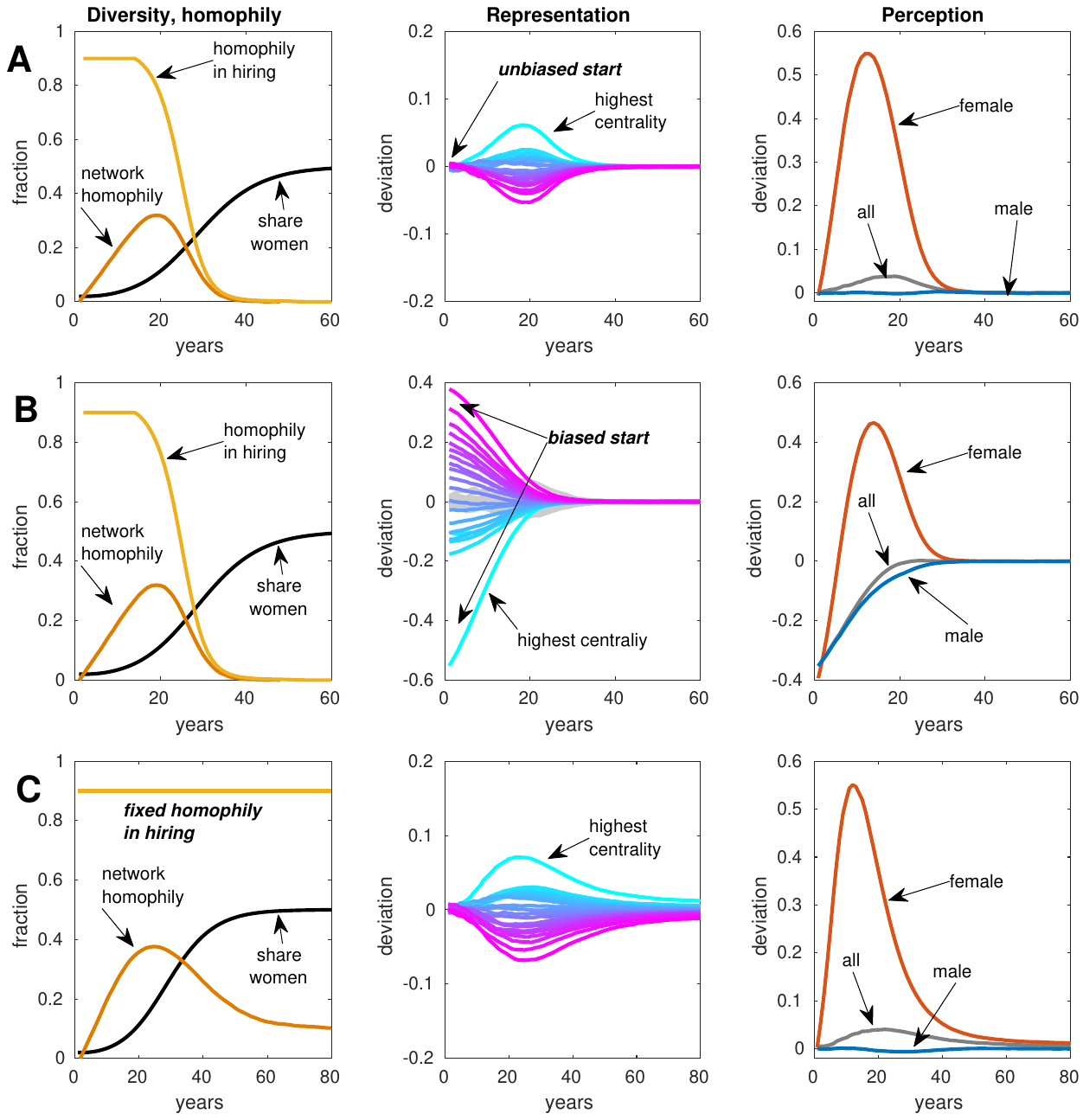}
  \end{center}
  \caption{Simulation results for the case of gender and three different scenarios. The left panels show the share of women in the network, as well as the homophily in hiring as well as the resulting homophily in the network. The middle panels show  the over/under-representation of women in firms with a certain centrality. The right panels show the perception of the group of women for female/male/all board members. A: baseline scenario, B: biased initial seat distribution, C: fixed homophily in hiring.}\label{fig:sim1}
\end{figure}

We will start with an analysis where the group attribute in our model can be interpreted as gender. The first scenario assumes that the initial assignment of seats is unbiased and that the homophily in hiring decreases with group size equalization. The results are shown as panels A in figure \ref{fig:sim1}.

The left panel shows that the homophily in hiring creates homophily in the firm network -- as can be expected -- but that this homophily also vanishes again over time. 
The middle panel shows the changes in the centrality of female board members -- defined by the centrality of the firms they work for. We show the over- and under-representation of women in 20 bins that represent the range of firm centralities from highest to lowest. We can see that the homophily in the hiring process creates a temporary boost in terms of women's average centrality, because the larger more connected firms are more likely to observe female board members in their neighborhoods, which increases the chance of hiring additional female board members. Hence, while in the generative network model introduced in section \ref{sec:models} the effect of homophily on the centrality of the minority is generally negative, we can show here that this is not necessarily the case for a diffusion process on such a network.

The right panel shows that homophily also influences the perception of women in local neighborhoods. Especially women themselves overestimate their own group size considerably, since they are likely to observe a much higher number of women in their neighborhoods than what can be observed in the entire population.

\subsubsection*{Biased seat assignments}

We can now change our assumption about the initial assignment of seats towards the biased scenario. This means that the initial assignment of seats  for women is inversely related to firm size (see equation \ref{eq:size}). The results are presented in panels B. We observe that a change in the initial seat assignment changes the resulting homophily in the network only marginally compared to the previous scenario (left panel). Instead of a temporary boost in perception, we observe a convergence towards equal representation in the long run (middle panel). The difference in the initial seat assignment has also implications for the perception of the group sizes (right panel). The bias towards board membership of women in small firms now leads to an initial underestimation of the group size of women for all other board members. This temporarily develops into of female over-estimation of their own size. For an illustration of the dependence of this development on $\gamma$, the amount of bias, the reader is referred to figure \ref{fig:vars} in the appendix.

Note that this scenario illustrates the possibility that the effects from behavioral homophily that we expect can be overshadowed by effects from the initial conditions for some time, despite them being present in the background.

\subsubsection*{Fixed homophily in hiring}

Finally, we are interested to find out if the vanishing of the outcome homophily requires a change in behavior during group equalization, or if it is an inherent feature that happens regardless of it. Therefore, we now assume that the homophily in hiring remains fixed, i.e., we permanently assign 90\% of board seats for women based on the number of women observed in the local neighborhoods, thus $\lambda(t) = \bar{\lambda}$.

The results are shown in panels C. Compared to the initial scenario (panels A) we can see that the boost in centrality (middle panel) is slightly increased, and the effects in perception are slightly prolonged. However, in both cases the effects almost vanish in the long run, despite the fixed homophily in hiring. This is due to the drop in homophily in the network (left panel), which does not completely disappear, yet drops to a level where its effects on representation and perception are very limited in the long run.
\begin{figure}[tb]
\begin{center}
    \includegraphics[width=0.8\textwidth, trim = 15 290 25 280, clip=true]{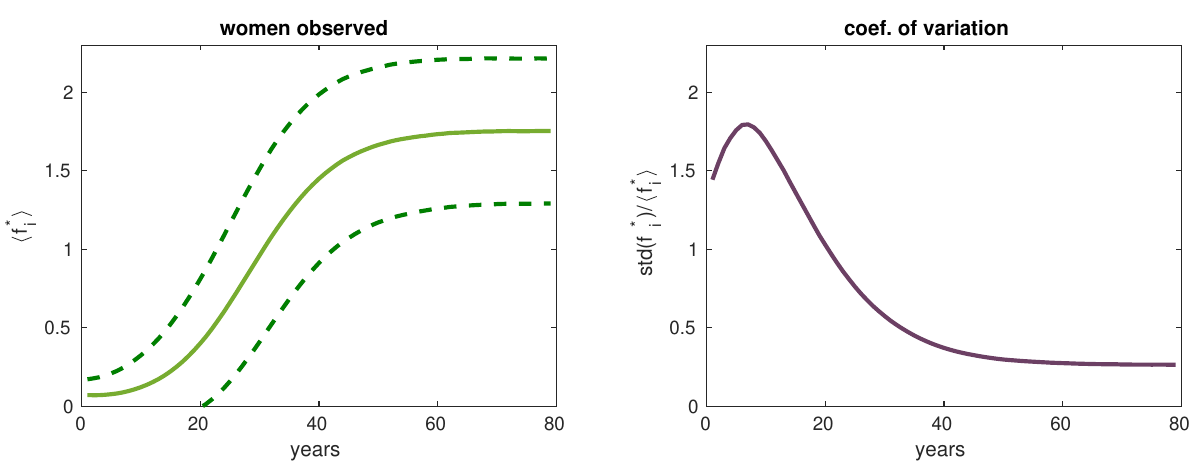}
  \end{center}
  \caption{Development of the average observed number of female board members $\langle f_i^* \rangle$ (and $\pm1$ standard deviation, left panel) and its coefficient of variation (right panel) for the scenario as shown in panels A of figure \ref{fig:sim1}. The coefficient of variation declines after an initial boom, signaling that the number of board seats with large differences with respect to their neighborhoods is shrinking over time, which shows the saturation effect in the network with respect to homophily.  }\label{fig:dif}
\end{figure}
Hence, there is a saturation effect in the network with respect to homophily. Even if the homophily in hiring is kept constant, the number of board seats with a homophilic neighborhood is limited.

We illustrate this effect by looking at the development of the variation in the number of observed female board members $f_i^* $, as described in by equation \ref{eq:fi}, and shown in figure \ref{fig:dif}. Since the average number of women is increasing over time, it is also necessary to look at the coefficient of variation, which is shown in the right panel. This can be interpreted as a measure of the level of homophily that can be achieved at each point in time, since it is an expression of the range of available neighborhoods with respect to gender. The homophily in the network follows this coefficient of variation (dark orange line in the bottom right panel of figure \ref{fig:sim1}) with some delay (because only 15\% of seats are re-assigned at each time step). The lower bound of the coefficient of variation is given by the natural noise of $f_i^*$ in neighborhoods with limited size. 

This result allows us to draw some conclusions for the question of the relationship between homophily and group size. It shows that diminishing homophily can -- at least in this case -- to a large part be explained as a structural phenomenon and requires little to no change of behavior of the individual. It thus illustrates how the intensity of the relationship between the homophily in behavior and the homophily as outcome in the firm network can change depending on the state of the system \citep[see also][]{saj_unveiling}.

\subsubsection*{Minorities}

In the previous scenarios we have assumed that the group sizes of men and women will equalize in the long run. When we drop this assumption, we arrive at a setting where the groups 
can be interpreted as a minority and the majority in the population. 
Simulation results that assume a minority which representation on corporate boards increases to its true size of $\frac{1}{6}$th of the population are shown in figure \ref{fig:sim2} in the appendix.

Most results are qualitatively close to those from the gender case. A difference is that a certain level of homophily is present in the network in the long run. This happens for two reasons. First, because the homophily in the hiring process does not completely vanish in the long run. This is, secondly, enabled by the fact that the network does not reach its saturation point when it comes to homophilic board seat assignments. The extent of these effects of course depends on the assumed homophily in hiring, in particular the functional form that describes its convergence.

\subsection{Scenarios with feedback from visibility}

\begin{figure}[tb]
\begin{center}
    \includegraphics[width=0.95\textwidth, trim = 5 410 10 130, clip=true]{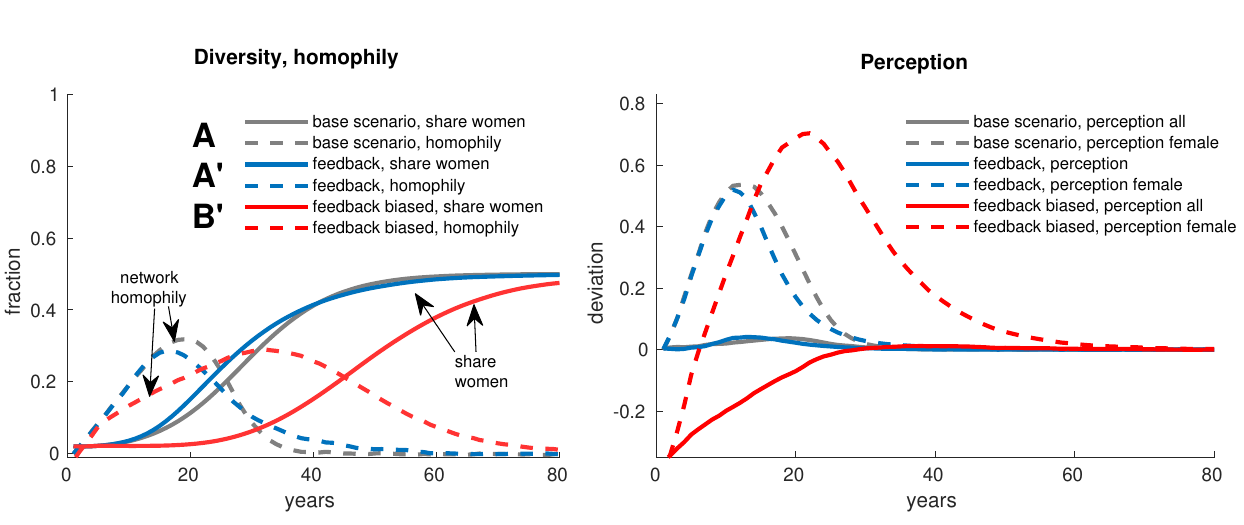}
  \end{center}
  \caption{Results for endogenous growth based on perception. The left panel shows the share of women (solid lines) and the corresponding network homophily (broken lines) for two scenarios with endogenous growth versus the base scenario corresponding to case A in figure \ref{fig:dif}. The right panel shows the bias in the perception of women for all board members (solid lines) and the perception bias of women with respect to their own group size (broken lines).}\label{fig:endo}
\end{figure}

Finally, we look at the case where the growth in the number of new board members is endogenous. Here we will focus on the case for gender.
The left panel of figure \ref{fig:endo} shows the change in the share of female board members over time. We can first compare the scenarios of an unbiased initial distribution of female board members with and without endogenous growth (scenario A versus A'). We observe that with endogenous growth (solid blue line) the increase in the number of female board members is slightly accelerated compared to the benchmark scenario (solid gray line). This is caused by the slight positive influence of perception bias, shown in the right panel (solid blue vs solid gray line). Note that the slightly quicker growth in the early phase is accompanied by an earlier peak of network homophily and perception itself, relative to the baseline scenario (broken lines in the left panel).

In the case of a biased start for women (towards smaller firms, scenario B') the influence of endogenous growth is more severe. This is the case, because the initial bias in centrality is accompanied by a significant negative perception bias (solid red line, right panel), which slows down the growth rate of  women on boards (solid red line, left panel).  This leads to a prolonged phase of significant network homophily and perception bias compared to both other scenarios.

Hence, an important feature of this network is the simultaneous bias in initial centrality and perception that has lasting consequences for the entire path of the minority towards equality. While it roughly takes two generations to achieve equality in our baseline model, including biases and assuming feedback means that one additional generation is needed to arrive at the same goal. While the precise timescale of this process is of course determined by the model calibration, this finding illustrates the importance of understanding visibility in networks. If we fail to identify and intervene to such biases early, we are wasting time in our efforts to realize diversity.

%\newpage
\section{Conclusions}\label{sec:conc}

%\Fariba{can we say something with regards to the effect of interventions and policies as oppose to the natural processes of ageing and replacement?}

The model presented in this paper can be viewed as an applied model of diffusion, social interaction, and imitation. Although stylized, it illustrates the influence of social and institutional structure on the dynamics of diversity. 

In particular, we show that homophily is not only a behavioral effect, but that its diminishing in the process of growth of a group is endogenous and thus structural. Interaction effects in the network can, therefore,  temporarily help minorities to grow into a more central position in the network. These effects depend heavily on the initial assignment of positions for the minority. The interdependence of homophily and visibility further plays a key role in the dynamics of diversity. Our model shows that once we allow for feedback from visibility, any bias in centrality will have devastating effects on the path to equality. 

This means that any intervention, for example, through a quota, should not only be discussed in terms of the overall target, but also in terms of its ability to remove biases with respect to visibility, and thus to enable a favorable growth path for minorities.

\subsection*{Declarations of interest} 
\noindent none

\printbibliography
\newpage
\section*{Appendix}

\begin{figure}[ht]
\begin{center}
    \includegraphics[width=0.85\textwidth, trim = 70 365 110 230, clip=true]{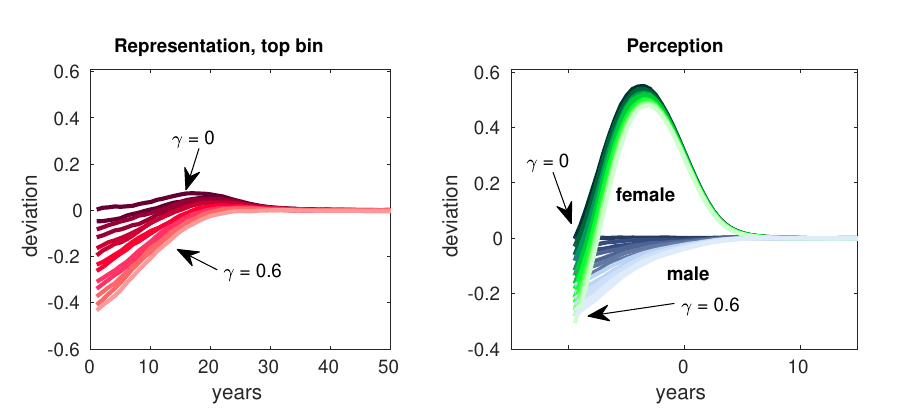}
  \end{center}
  \caption{Representation in terms of eigencentrality and perception for different levels of bias in the initial assignment of seats. $\gamma$ is varied in 13 steps from $0$ to $0.6$. The left panel shows the over/under-representation of women in the top-5\%-bin. The right panel shows the perception of the group size of female board members for male and female board members.}\label{fig:vars}
\end{figure}

\begin{figure}[htb]
\begin{center}
    \includegraphics[width=\textwidth, trim = 15 370 10 30, clip=true]{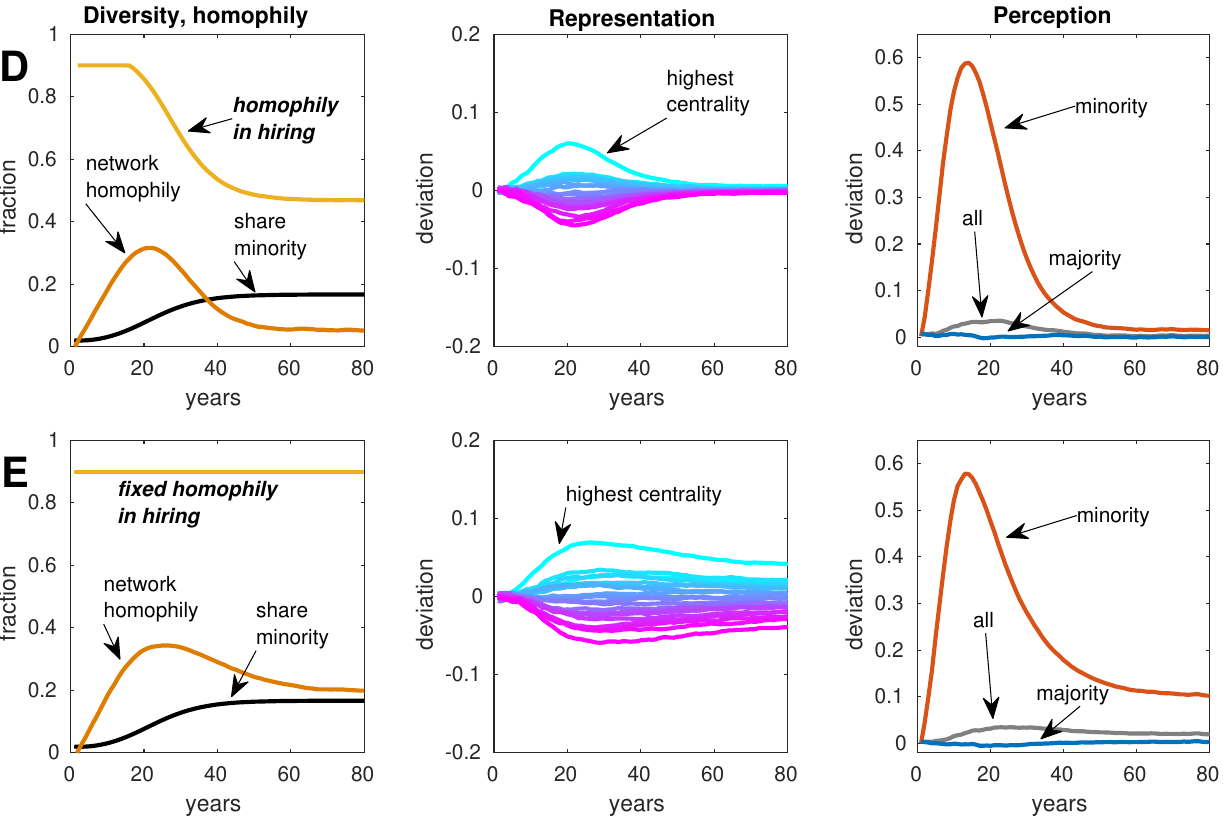}
  \end{center}
  \caption{Simulation results for the case of a minority and two different scenarios. The left panels show the share of the minority in the network, as well as the homophily in hiring as well as the resulting homophily in the network. The middle panels show  the over/under-representation of the minority in firms with a certain centrality. The right panels shows the perception of the minority for minority/majority/all board members. D: baseline scenario, E: fixed homophily in hiring.
In terms of representation (middle) panel we observe that homophily can for minorities actually lead to measurable effects in terms of centrality in the long run, as well as lasting effects in group perception (right panel). 
  }\label{fig:sim2}
\end{figure}

\end{document}